\documentclass[a4paper,10pt]{article}
\pdfoutput=1 

\usepackage{jinstpub} 

\usepackage[utf8]{inputenc}
\graphicspath{{./}}
\usepackage{microtype}                
\usepackage{cleveref}                 
\usepackage{eurosym}                  
\usepackage[binary-units]{siunitx}    
\usepackage{mhchem}                   
\usepackage{textgreek}                
\usepackage{amsfonts,stackrel,amsthm} 
\usepackage{color} 

\usepackage{tikz}
\usetikzlibrary{shapes.geometric,arrows,positioning}
\usetikzlibrary{fadings,backgrounds,fit}

\usepackage{shellesc}                 
\usetikzlibrary{external}             
\definecolor{ippBlue}{RGB}{0, 101, 191}
\definecolor{ccfeOrange}{RGB}{243, 148, 0}
\definecolor{mpGreen}{RGB}{0, 123, 108}
\definecolor{taRed}{RGB}{217,41,59}

\tikzset{
  treenode/.style   = {shape=rectangle,
    draw=black, very thick, text=white, fill=ippBlue,
    inner sep=1mm, minimum height = 15mm},
  mainDSP/.style    = {treenode},
  supportDSP/.style = {treenode, fill = ccfeOrange},
  extDSP/.style = {treenode, fill = taRed},
  input/.style      = {treenode,
    minimum height=6mm, minimum width=6mm},
  output/.style     = {treenode,
    minimum height = 6mm, minimum width=6mm},
  arithOp/.style      = {supportDSP, minimum height = 4 mm,
    shape=circle},
  signal/.style     = {very thick}, 
  dirSignal/.style  = {signal, ->, >=stealth'},
  multiSignal/.style  = {dirSignal, double},
  2waySignal/.style = {signal, <->, >=stealth'},
  optSignal/.style   = {line width = 3pt, color = mpGreen,
  ->, >=stealth'},
}

\newcommand{\includepdf}[1]{\includegraphics[]{./#1.pdf}}

\title{Compensation of phase drifts caused by ambient humidity, temperature and pressure changes for continuously operating interferometers}

\author[a,1]{K. J. Brunner,\note{Corresponding author.}}
\author[a]{J. Knauer,}
\author[a]{J. Meineke,}
\author[a]{M. Stern,}
\author[a]{M. Hirsch,}
\author[a]{B. Kursinski,}
\author[a]{R. C. Wolf}
\author[2]{and the W7-X team\note{for a full list of contributors, see Klinger \emph{et al.}, 2019\cite{Klinger2019}}}

\affiliation[a]{Max-Planck-Institute for Plasma Physics, Wendelsteinstr. 1, 17489 Greifswald, Germany}

\emailAdd{k.j.brunner@ipp.mpg.de}

\abstract{Fusion experiments rely heavily on the measurement of the line-integrated electron density by interferometry for density feed-back control. In recent years the discharge length has increased dramatically and is continuing to rise, resulting in environmentally induced phase drifts to become an increasingly worrisome subject, since they falsify the interferometer's measurement of the density. Especially in larger Tokamaks the loss of density control due to uncontrolled changes in the optical path length can have a disastrous outcome. The control of environmental parameters in large diagnostic/experimental halls is costly and sometimes infeasible and in some cases cannot be retro-fitted to an existing machine. In this report we present a very cheap (ca.~\EUR{100}), easily retro-fitted, real-time capable phase compensation scheme for interferometers measuring dispersive media over long time scales. The method is not limited to fusion, but can be applied to any continuously measuring interferometer measuring a dispersive medium. It has been successfully applied to the Wendelstein 7-X density feed-back interferometer.}


\def\acknowledgement{\itshape This work has been carried out within the framework of the EUROfusion Consortium and has received funding from the EURATOM research and training programme 2014-2018 and 2019-2020 under grant agreement No 633053. The views and opinions expressed herein do not necessarily reflect those of the European Commission.}

\usepackage{layouts}
\begin{document}

\maketitle
\flushbottom 

\section{Introduction}
\label{sec:intro}

The discharge length of fusion machines so far has been of the order of a few minutes only. However, as fusion research approaches reactor relevant conditions the discharge lengths increase. Fusion machines such as Wendelstein 7-X (W7-X) aim at discharge lengths of \SI{15}{\minute} at full power\cite{Klinger2019}. ITER, the first plasma experiment to reach ignition, will run a base-line length of \SI{30}{\minute} shots and the current European demonstration power plant (DEMO), will run at least on basis of a \SI{2}{\hour} cycle\cite{Snipes2012,Biel2019}. Already machines such as QUEST have achieved such extensive discharges\cite{Hanada2017}.

Interferometry is the primary density control diagnostic for fusion machines in the world. All of the aforementioned machines have or will have continuous interferometry based real-time density control, albeit the real-time requirements differ wildly between these machines\cite{vanZeeland2013,Biel2019}. The techniques primary advantage is its simplicity. To first order it is only sensitive to the dispersion the traversed medium and the wavelength of the employed laser beam. And although vibrations do affect the measured phase, advanced techniques such as dispersion interferometry or the well-established two-color interferometry have successfully eliminated their effect\cite{Mlynek2010,Boboc2012,Akiyama2014,vanZeeland2017,Brunner2018}.

However, what has thus far been neglected is the effect of environmental parameters, i.e. air-humidity, air-temperature and air-pressure, on the phase measurement. The reason is that for most applications to date the environmental parameters are either considered constant over the course of the measurement, or (as in the case of the LIGO system) have been mitigated by evacuating the entire optical setup. For specific components, such as \ce{ZnSe}-vacuum windows, temperature-induced phase-drifts have been presented at TJII and post-processing correction methods were consequently developed\cite{Sanchez2005}. The phase-drifts presented were the result of absorbed microwave stray-radiation and could be easily mitigated by choice of an appropriate wavelengths-combination. The overall environmental parameters were not considered. 

In general an interferometer placed at a fusion experiment will have the primary optical components outside of the test cell far away from the point of measurement, which advantageously allows the diagnostic to be maintained while the fusion experiment continues to run. This in turn requires large parts of the optical beam path to be in air. For \mbox{DIII-D}, which is the test-bed for the ITER interferometer, the beam path though air is of the order of \SI{120}{\m}\cite{vanZeeland2017}. For DEMO this will potentially increase by a factor of 2.

\begin{figure}[t]
  \centering
  \includepdf{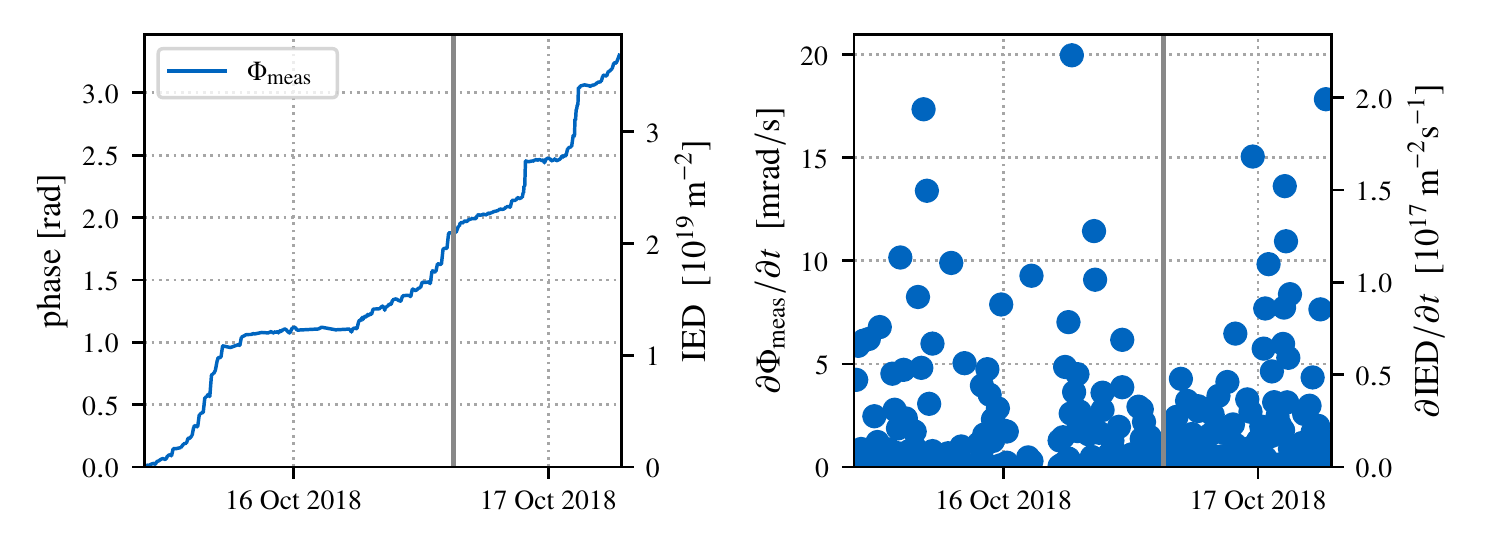}
  \caption{\label{fig:motivation} Measured phase drift for Oct 16 \& 17, 2018 of the OP1.2b operation campaign at Wendelstein 7-X. The actual drifting phase is shown in the left-hand plot. The ordinates indicate the phase value on the left and the corresponding IED equivalent for the W7-X system on the right. The right-hand plot shows the rate of phase drift for this day. For visualization purposes the abscissa is not to scale. The gray line indicates the day break.}
\end{figure}

It has recently come to the attention of the fusion community that the diagnostic-hall climate has a profound impact on the phase measurement of fusion interferometers\cite{vanZeeland2017}. In particular air humidity was shown to have an impact on the measurement of the phase\cite{Brunner2018}. \Cref{fig:motivation} shows the phase measured by the W7-X interferometer on Oct 16 \& 17, 2018, which was part of the OP1.2b operation campaign. The left-hand plot shows the plain phase with the equivalent line-Integrated Electron Density (IED) indicated on the right ordinate. As can be seen a changing weather front resulted in significant phase drifts over the course of the day. On the right plot the rate of phase-drift is indicated, where the box-averaging was used to remove short-scale drifts, i.e. only the drifts from discharge to discharge are shown. As can be seen the phase drifts can (under bad circumstances) amount to \SI{\approx1e17}{\m^{-2}\s^{-1}}. The maximum shot length during the OP1.2b operation campaign was already \SI{100}{\s} yielding a maximum drift-induced density error of \SI{1e19}{\m^{-2}}. The projected maximum shot length of \SI{\approx1000}{\s} would correspondingly result in a \SI{1e20}{\m^{-2}} density error, which would be unacceptable. 

The data in \cref{fig:motivation} shows, that phase drifts must be compensated for semi-continuously operating interferometers. In this paper we present a real-time compensation method to stabilize the phase measurement of a long base-line interferometer for the purpose of continuous density feed-back control. The system is very easily retro-fitted to already existing interferometers and can be implemented with a very low budget. The system was implemented at W7-X as part of the single channel dispersion interferometer\cite{Knauer2016,Brunner2018}.

In \cref{sec:model} the compensation model will be derived. Its implementation in the W7-X interferometer system will be detailed in \cref{sec:impcalib}, where the method of calibration is also described. The effectiveness of the compensation will be revealed in \cref{sec:results}, followed by a discussion and an outlook in \cref{sec:discussion}.

\section{Environmental Phase Model}
\label{sec:model}

The phase measured by an interferometer is generally the combination of various contributions. This is because an interferometer is sensitive to both the physical path length difference as well as the refractive index of the traversed medium (the combination of both is known as the optical path length). Generally, the contribution of interest is only a part of the measured phase.

\begin{equation}
  \label{eq:phase}
  \Phi_{\text{meas}} = \Phi_{\Delta L} + \Phi_{\text{disp. med.}} + \Phi_{\text{interest}} + \delta\Phi 
\end{equation}
    
For the application of nuclear fusion the interesting quantity is the dispersion of the fusion plasma $\Phi_{\text{interest}}=\Phi_{\text{plasma}}$. The perturbing quantities in this instance are the difference in path length $\Phi_{\Delta L}$ as well as the change in refractive index of the dispersive components in the optical setup $\Phi_{\text{disp. media}}$. In \cref{eq:phase} $\delta\Phi$ is the contribution of phase errors related to the diode signal evaluation. 

In large-scale experiments $\Phi_{\Delta L}$, which are ubiquitous in fusion, will generally dominate, unless the interferometer has extremely long wavelengths. These are however unfavorable for high-performance fusion machines, since they result in high levels of refraction. As such, fusion experiments rely on interferometric measurements such as two-color interferometry (2CI) or dispersion interferometry (DI) to remove the $\Phi_{\Delta L}$-contribution\cite{Brunner2015,Brunner2018}.With appropriately calibrated systems $\Phi_{\Delta L}$ can be neglected. Ignoring errors from the signal evaluation, the measured phase in a fusion setting is therefore :

\begin{equation}
  \label{eq:dispPhase}
  \begin{split}
    \Phi_{\text{meas}} &= \Phi_{\text{beam path}} + \Phi_{\text{plasma}} = \Phi_{\text{trans. comp.}} + \Phi_{\text{air}} + \Phi_{\text{plasma}} \\
    &\approx \frac{L_{\text{air}}(T, p, H)}{\lambda} N_{\text{air}} (\lambda, T, p, H) + \sum_c \frac{L_{c}(T, p, H)}{\lambda} N_c (\lambda, T, p, H) + \Phi_{\text{plasma}} \\
    &= \lambda^{-1} \left<L N\right>(\lambda, T, p, H)  + \Phi_{\text{plasma}}.
  \end{split}
\end{equation}

In most fusion settings the optical interferometer setup is installed in areas, where environmental conditions cannot be tightly regulated. As such, the dispersive disturbance is generally the sum of contributions from the transmission components '$c$', e.g. lenses, as well as the refractive index of air (see \cref{eq:dispPhase}). Without a loss of precision, the phase contribution due to the optical path length of the individual components can be equivalently expressed by a single \emph{average} optical path length $\left<L N\right>$. Note that this quantity is a function of the environmental parameters, i.e. the temperature $T$, the air pressure $p$ and the (absolute) humidity $H$, as well as the wavelength $\lambda$ of the traversing light. For this reason, the measured phase tends to drift over time. Due to the long path lengths in large scale experiments, $\Phi_{\text{air}}$ will tend to dominate and has been shown to be a significant contribution to the measured phase\cite{Brunner2018}. 

R.~J.~Mathar pointed out that the refractive index of air can be approximated by a Taylor series\cite{Mathar2007}. It was also shown at W7-X that the approximation can be reduced to only the 0th order component of eqn.~(5)~\&~(6) of Mathar's approximation and still produce an acceptable fit\cite{Brunner2018}. While this approximation is published for air only, it is not unreasonable to assume that a similar approximation is applicable to any dispersive transmission component in the beam path of the laser, e.g. lenses. Since the approximation describes the refractive index as a polynomial, one can therefore sum the individual contributions of each component to yield one approximation for the mean optical path length $\left<L N\right>$ in \cref{eq:dispPhase}. 

\begin{equation}
  \label{eq:refrIdxFourier}
  \begin{split}
    \left<L N\right>(\lambda, T, p, H) \approx & s_0 - s_{H2}*H_{\text{air}}^2 - s_{H1}*H_{\text{air}} - \\
    & s_{p2}*p_{\text{air}}^2 - s_{p1}*p_{\text{air}} - s_{T2}*T_{\text{air}}^2 - s_{T1}*T_{\text{air}} - \\
    & s_{\text{Tp}}*T_{\text{air}}*p_{\text{air}} - s_{\text{TH}}*T_{\text{air}}*H_{\text{air}} - s_{\text{pH}}*p_{\text{air}}*H_{\text{air}}.
  \end{split}
\end{equation}

In \cref{eq:refrIdxFourier} we have summed the Fourier coefficients of each component to form a single polynomial, i.e. $s_n \approx L_{\text{air}} c_{n,\text{air}}(\lambda) + \sum_{\text{comp.}} L_{\text{comp.}} c_{n,\text{comp}}(\lambda)$. This approximation assumes that the transmission component has some proportionality relation of its refractive index with the ambient environmental parameters. This is reasonable as each component will for example equilibrate its internal temperature (which can be far from $T_{\text{air}}$) according to the ambient air temperature by convection. The coefficients also deal with the relative length variations due to thermal expansion. All \emph{constant} factors are summed up in $s_0$.

In general the measured phase $\Phi_{\text{meas}}$ will be a function of the sum of \cref{eq:refrIdxFourier} for two wavelengths $\lambda$. However, one can make use of the fact, that the ratio of these two wavelength is fixed, i.e. $\lambda_1 = \text{\emph{const.}} \cdot  \lambda_2$. For a DI that constant is 2 exactly, but even for a 2CI the ratio is fixed. Given this circumstance the difference of the Fourier constants $s_n$ in \cref{eq:refrIdxFourier}, which are only a function of the wavelength, can be simplified into a single constant by

\begin{equation}
  \label{eq:lambdaScale}
  s_n(\lambda_1) - s_n(\lambda_2) = s_n(\lambda_1) - s_n(A\cdot\lambda_1) = s_n(\lambda_1) - B\cdot s_n(\lambda_1) \equiv e_n(\lambda_1)
\end{equation}

Combining \cref{eq:dispPhase,eq:refrIdxFourier,eq:lambdaScale} therefore yields a comparatively simple equation to remove the environmentally induced phase drift from an interferometer measurement.

\begin{equation}
  \label{eq:corrEqn}
  \begin{split}
    \Phi_{\text{plasma}} = \Phi_{\text{meas}} - (&e_0 + e_{H2}*H_{\text{air}}^2 + e_{H1}*H_{\text{air}} + \\
    & e_{p2}*p_{\text{air}}^2 + e_{p1}*p_{\text{air}} + e_{T2}*T_{\text{air}}^2 + e_{T1}*T_{\text{air}} + \\
    & e_{\text{Tp}}*T_{\text{air}}*p_{\text{air}} + e_{\text{TH}}*T_{\text{air}}*H_{\text{air}} + e_{\text{pH}}*p_{\text{air}}*H_{\text{air}} )
  \end{split}
\end{equation}

Note that at this point the $e_n$ are unknown constants, which depend on the laser's wavelength (for 2CI an arbitrary choice may be made) and the non-evacuated beam path length. 

\section{Implementation \& Calibration}
\label{sec:impcalib}

Since the $e_n$ in \cref{eq:corrEqn} are usually given by the design of the system and do not change (significantly), once it has been taken into operation, they can generally be assumed constant for a given interferometer system. As such it is possible to find them by correlating measurements of environmental parameters ($H$, $p$ and $T$) with the measurements of the measured phase $\Phi_{\text{meas}}$ in absence of a plasma induced phase shift, i.e. $\Phi_{\text{plasma}} = 0$. This measurement was conducted with the W7-X integral electron density dispersion interferometer (IEDDI) described in the next section. Insufficient air conditioning during the 2018 operation campaign (OP1.2b) at W7-X enabled the calibration measurement, as shown in \cref{ssec:calibration}.

\subsection{Implementation in the W7-X IEDDI system}
\label{ssec:implementation}

The W7-X IEDDI system is modulated dispersion interferometer utilizing a \SI{10.6}{\um} \ce{CO2} laser for line-integrated density measurements as first developed at TEXTOR\cite{Bagryansky2006}. The system uses a novel real-time phase extraction method for the diode signal, which is based on a field programmable gate array (FPGA) signal processor\cite{Brunner2018}. It samples the diode signal at \SI{50}{\MHz} and conducts real-time processing of the diode signal in under \SI{40}{\us}. The total beam path is around \SI{50}{\m}, which is predominantly placed on an optical table next to the stellarator in the W7-X torus hall\cite[fig.2]{Knauer2016}. This circumstance was actually advantageous for the development of this method, as it can be assumed that the majority of the beam path is governed by the same environmental parameters.

\begin{figure}[t]
  \centering
  \includepdf{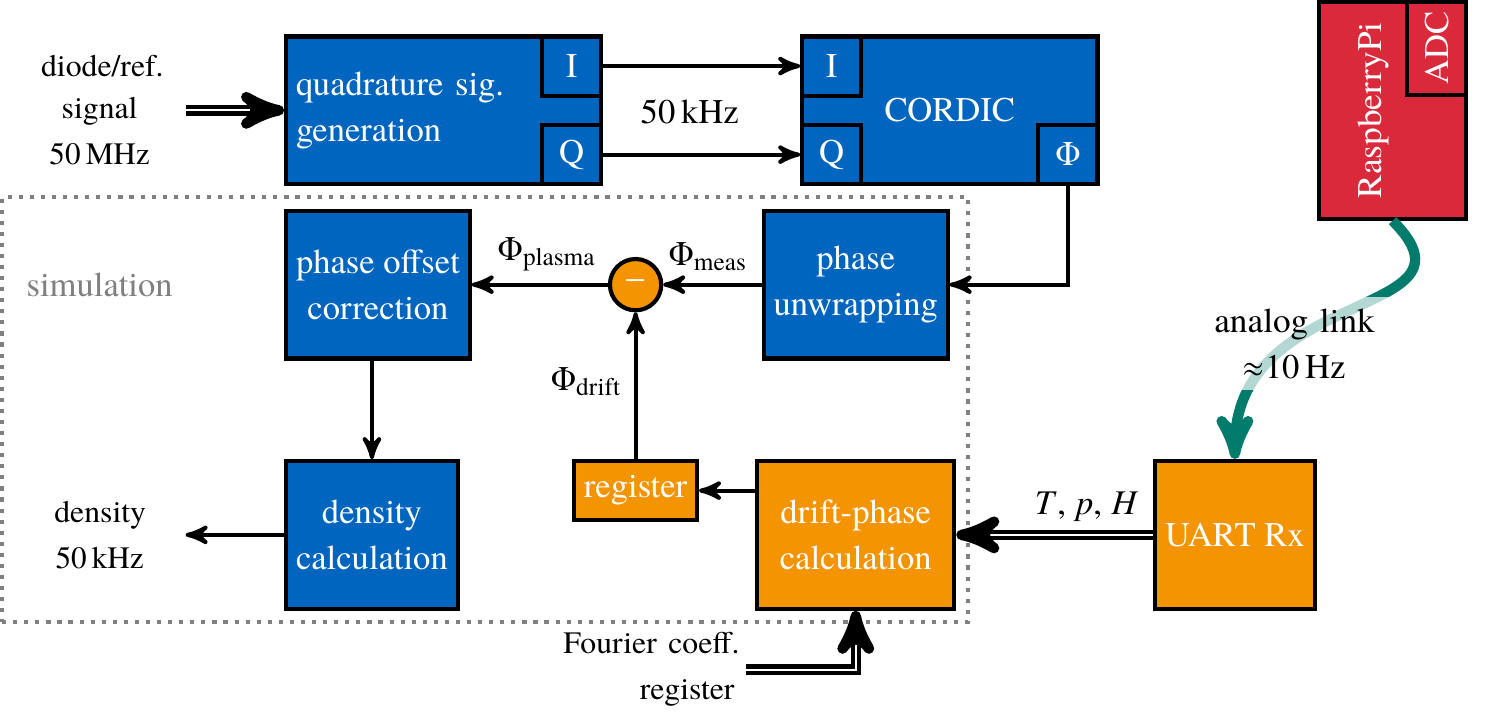}
  \caption{\label{fig:DSPtop} The structure of the firmware's DSP core. The original implementation is indicated in blue and has been detailed previously\cite{Brunner2018}. The external environmental sensor is indicated in red. The logic applying the phase drift model detailed in \cref{sec:model} to the real-time phase is indicated in orange. The logic simulated in \cref{sec:results} is marked.}
\end{figure}

\Cref{fig:DSPtop} shows the structure of the real-time phase evaluation firmware. The blue cores are the original implementation described in an earlier publication\cite{Brunner2018}, whereas the modifications to the FPGA firmware are depicted in orange. Before the start of the OP1.2b campaign the system was equipped with an environmental sensor based on a Raspberry~Pi~3B+ combined with a SenseHAT (shown in red in \cref{fig:DSPtop})\cite{raspberryPi}. It was placed at the center of the optical table, measuring continuously. The Rasperry~Pi measures temperature, pressure and relative humidity on the time scale of the sensor, which is roughly every \SI{100}{\ms}. The data is written to the central W7-X data storage using Ethernet and in parallel sent via a co-axial cable to the real-time processing FPGA using a universal asynchronous receiver transceiver (UART) protocol. 

The observant reader will have noticed that \cref{eq:corrEqn} uses the absolute humidity, while the SenseHAT measures relative humidity. This does not pose a problem, since the relative humidity can be related to the absolute humidity using the perfect gas law and the Arden Buck equation\cite{Buck1981}. Using a basic Fourier expansion again, the deviation are additional temperature and pressure terms, which would simply add themselves to the coefficients in \cref{eq:corrEqn}, merely changing the actual value of the $e_n$.

The FPGA firmware was fitted with a UART frame receiver. It translates the data received from the Raspberry~Pi into a format understood by the FPGA firmware. The measurement of temperature, rel. humidity and pressure are then passed to a drift prediction core, which calculates a drift phase $\Phi_{\text{drift}}$ based on the polynomial inside the parentheses of \cref{eq:corrEqn}. The final correction is then conducted by basic subtraction of $\Phi_{\text{drift}}$ from the plain measured phase $\Phi_{\text{meas}}$. This is the only modification to the previous data flow in the firmware. This is highly beneficial as it demonstrates the ease with which the scheme can be retro-fitted to existing systems. 

The drift-phase $\Phi_{\text{drift}}$ is calculated every time new samples of environmental parameters arrive at the FPGA (roughly every \SI{100}{\ms}). A register is used to transfer the slow stream to the processed stream. There is a significant delay between measuring the environmental parameters at the Rasperry~Pi their ``application'' on the FPGA. However, this time-lag should be negligible compared to the time-scales on which the environmental parameters change. The Fourier coefficients for the correction are fixed in registers, which can be set externally. This can even be done continuously during the operation.

\subsection{Calibration of Fourier Coefficients}
\label{ssec:calibration}

Calibrating the Fourier coefficients in \cref{eq:corrEqn} is the primary challenge when implementing this method. Since (at W7-X) the environmental parameters in the torus hall (TH) cannot be controlled arbitrarily, it was necessary to rely on the environmental parameters changing naturally, e.g. due to changing weather fronts. This inherently resulted in a very long time to conduct the calibration measurements. The necessary measurements took the entire 2018 operation campaign (OP1.2b). Humidity and temperature varied significantly during this time due to the very hot summer close to the sea with changing weather fronts passing through Greifswald.

The calibration measurements have to be conducted through the beam path that requires the correction, e.g. torus vessel. Safety measures prevented a continuous measurement, since the laser's shutters have to be closed when people were enter the W7-X TH. Therefore, the necessary data had to be taken from the interferometer's offset measurements. More explicitly the raw data acquired by the interferometer before the onset of plasma heating was evaluated according to the scheme depicted in \cref{fig:DSPtop}. This yielded $\Phi_{\text{meas}}$ \emph{without} the offset correction. A final continuous measurement over 4 consecutive days was recorded at the end of the campaign to complete the data set. 

\begin{figure}[t]
  \centering
  \includepdf{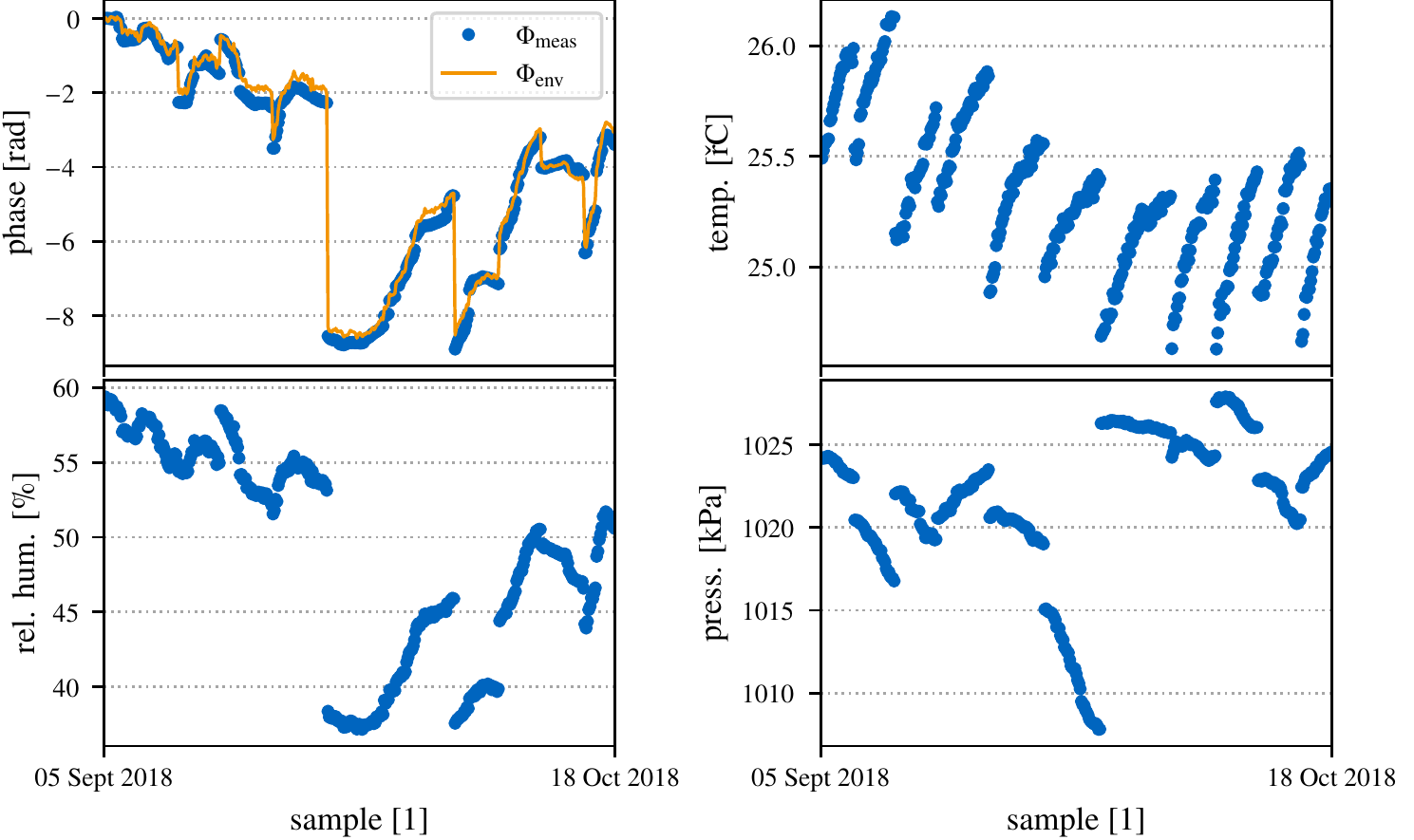}
  \caption{\label{fig:envCalib} Environmental drift calibration measurement during the OP1.2b operation campaign. Each data point plotted is a \SI{1}{\s} box average of the values recorded during the time. To better visualize the data, the abscissa is not to scale, and the data has been decimated. The fitted phase model of \cref{eq:corrEqn} is plotted in orange on the top left. The data of Oct 16 \& 17, 2018 have been excluded from the graph and the fit to prevent falsification of the simulation presented in \cref{sec:results}.}
\end{figure}

\Cref{fig:envCalib} shows the calibration measurement and subsequent model fit. The temperature, air pressure and \emph{relative} humidity, as measured by the Raspberry~Pi are shown on the top right, bottom right and bottom left respectively. The measured phase is shown on the top left in blue and the fitted drift phase $\Phi_{\text{drift}}$ according to \cref{eq:corrEqn} in orange. Each data point is a \SI{1}{\s} box average of all data recorded. Since time is not a fitted quantity the natural spaces between each point have been omitted, i.e. the abscissa is not to scale.

Even before considering the fitted curve, the data reveals a strong correlation between relative humidity and the measured phase. It is also obvious that the optical table of the W7-X system heats up over the course of an operation day. Quantifying the correlation is however more difficult. The most applicable correlation metric is the distance correlation $\mathcal{R}$, which enables a correlation metric for non-linear dependencies\cite{Szekely2007}. However, due to the coupling between the model parameters ($T$, $p$ and $H$), ``plain'' distance correlation does not deliver a good metric of the individual dependencies. The ``standard'' approach to remove the unwanted combined coupling is partial correlation, which removes the effect of a third variable coupled to the two variables of interest\cite{Szekely2014}. Since there are always two ``confounding'' variables for any combination of environmental parameters and the phase, the average partial distance correlation $\bar{\mathcal{R}}(\Phi;H|p,T) = 0.5 \cdot \left( \mathcal{R}(\Phi;H|p) + \mathcal{R}(\Phi;H|T) \right)$ for any permutation of $p$, $T$ and $H$ is calculated to yield an indicator for the level of correlation. Given the data-set in \cref{fig:envCalib} the partial distance correlation factors are as follows :

\begin{itemize}
  \centering
  \item[$\mathcal{R}(\Phi;H|T,p)$] : 0.986
  \item[$\mathcal{R}(\Phi;T|H,p)$] : 0.443
  \item[$\mathcal{R}(\Phi;p|H,T)$] : 0.187
\end{itemize}

As can be seen there is a very high level of correlation between  the humidity and phase, whereas the correlation between pressure and phase is relatively small. This emphasizes, that the humidity is the primary driving factor for the phase drifts of the W7-X interferometer. The fitted drift phase in orange traces the measured phase very well. The errors are within the \SI{0.6}{\radian} natural phase extraction error of the phase evaluation technique used by the W7-X system\cite{Brunner2018}. 

\section{Results}
\label{sec:results}

To show the effectiveness of the calibration derived in the previous section the real-time correction must be shown using the FPGA real-time evaluation. Unfortunately, since the W7-X interferometer as a control diagnostic had to be fully available for the operation campaign, it was not possible to test the correction already during the course of the OP1.2b campaign. Nonetheless, the firmware was appropriately modified. For the operation and the correction module simply ``disabled''. Fortunately, the benefit of an FPGA logic design is that it is well-behaved and can be simulated. The W7-X system has the additional benefit of storing most of its raw data, i.e. the data that is fed into the real-time evaluation logic of the FPGA, to support system-developments.

It was therefore possible to demonstrate the effectiveness of the compensation presented here directly using a logic simulation. Since the simulation is computationally very expensive a compiled logic simulation based on the open-source tool GHDL by Tristan Gingold was written\cite{ghdl}. To further reduce the computational effort required, the simulation only included the logic relevant to the compensation scheme presented here as indicated by the box in \cref{fig:DSPtop}.

The logic test bench written simulated the acquisition of an interferometer shot exactly as it happens during a W7-X discharge\cite{ghdl}. The primary simulation logic is indicated in blue and orange in \cref{fig:DSPtop}. As inputs to the system raw data acquired by the FPGA during normal operation. No averaging was conducted. Instead the raw diode and reference signal of a full period of a dispersion interferometer modulation was taken every \SI{100}{\ms}. From this data the wrapping phase (as supplied by the CORDIC core in \cref{fig:DSPtop}) was calculated. The wrapping phase was then fed into the simulation at the appropriate point. The environmental parameters were chosen at a lower rate and spaced randomly to appropriately mimic the system behavior. While this is not exactly what the FPGA would see during an actual operation, it shows that the model works, since phase evaluation and drift calculation are independent of each other and do not depend on previous modulation periods.

\begin{figure}[t]
  \centering
  \includepdf{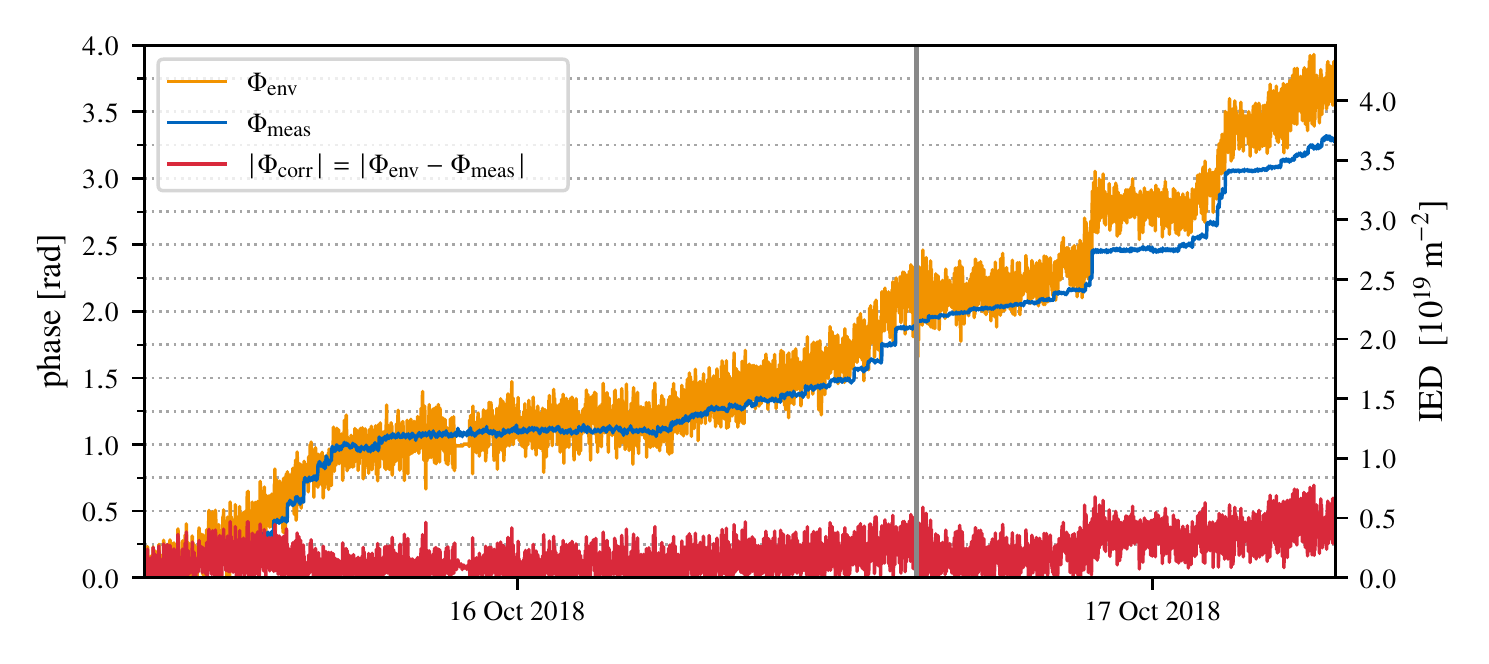}
  \caption{\label{fig:FPGAtest} Demonstration of phase drift stabilization using a GHDL simulation. The simulation input data is real data from Oct 16 \& 17, 2018. The measured drifting phase is shown in blue and the modeled drift phase in orange. The absolute value of the compensated phase is shown in red. The left ordinate indicates the phase in radian and the right one the equivalent error to the IED in \SI{e19}{m^{-2}}. For visualization purposes the abscissa is not to scale. The gray line indicates the day break.}
\end{figure}

\Cref{fig:FPGAtest} shows the results of the simulation. The simulation data was taken from Oct 16 \& 17, 2018, i.e. during the last operational week of the OP1.2b campaign and the same data shown in \cref{fig:motivation}. Note that this data was excluded from the fit in \cref{ssec:calibration}, as to not distort the result. The plot depicts the drifting measured phase $\Phi_{\text{meas}}$  in blue with the modeled phase $\Phi_{\text{env}}$  in orange. The compensated phase $\Phi_{\text{corr}}$ is shown in red, which corresponds to $\Phi_{\text{plasma}}$ in \cref{fig:DSPtop} and \cref{eq:corrEqn}. The equivalent density error is marked on the right ordinate. As can be seen, the phase drift is reduced by an order of magnitude, yielding a compensated phase error of only \SI{0.4}{\radian} or and IED of \SI{\approx4e18}{m^{-2}} over the course of 2 days, during which the measured phase drifted significantly.

The plot shows a continuously increasing gap between the modeled phase and the measured one. However as has been noted before, there is an error associated with the compensation, which is of the order of the natural phase error of the W7-X phase evaluation algorithm. There is a high probability that the gap is due to this error. Addressing these errors will be topic of other publications.

An additional issue obvious from \cref{fig:FPGAtest} is the increase ins statistical noise of the compensated signal. It is evident that the source is the fit model $\Phi_{\text{env}}$, which in turn is subject to the noise in the environmental parameter measurement. 

\section{Discussion \& Outlook}
\label{sec:discussion}

We have shown that it is possible to reduce the phase drift induced into dispersion measuring interferometers by a simple measurement of air temperature, humidity and pressure using cheap hardware based on a Raspberry~Pi. The total hardware cost of implementing this compensation scheme was around \EUR{100}. While the system could not be shown to operate in-situ, logic simulation with data taken during the actual operation of the system show a reduction of the phase drift by an order of magnitude. 

The primary difficulty is calibration of the system, which requires a lengthy measurement of environmental parameters. However, the improvement of the calibration can be conducted continuously by recording the measured phase and improving the Fourier coefficients over time. The compensation method presented here is not specific to fusion, although this appears to be one of the fields, were long time scale phase drifts are a prominent problem. Nonetheless any interferometer conducting a similar measurement scheme could implement this, e.g. measuring the dispersion of a material like water over long periods of time. 

The measurement presented here was conducted with a relatively short optical beam path of "only" \SI{50}{\m}. Larger systems such as ITER and DEMO will have optical beam path of more than \SI{100}{\m}. It is foreseeable, that a single measurement of environmental parameters does not suffice to conduct the appropriate fit. However, one can simply split \cref{eq:corrEqn} into several sub-paths of L. This results in a sum of two polynomials with two different sets of coefficients. Nonetheless the fitting procedure would not change. This must be tested on an appropriate set-up. 

\begin{figure}[t]
  \centering
  \includepdf{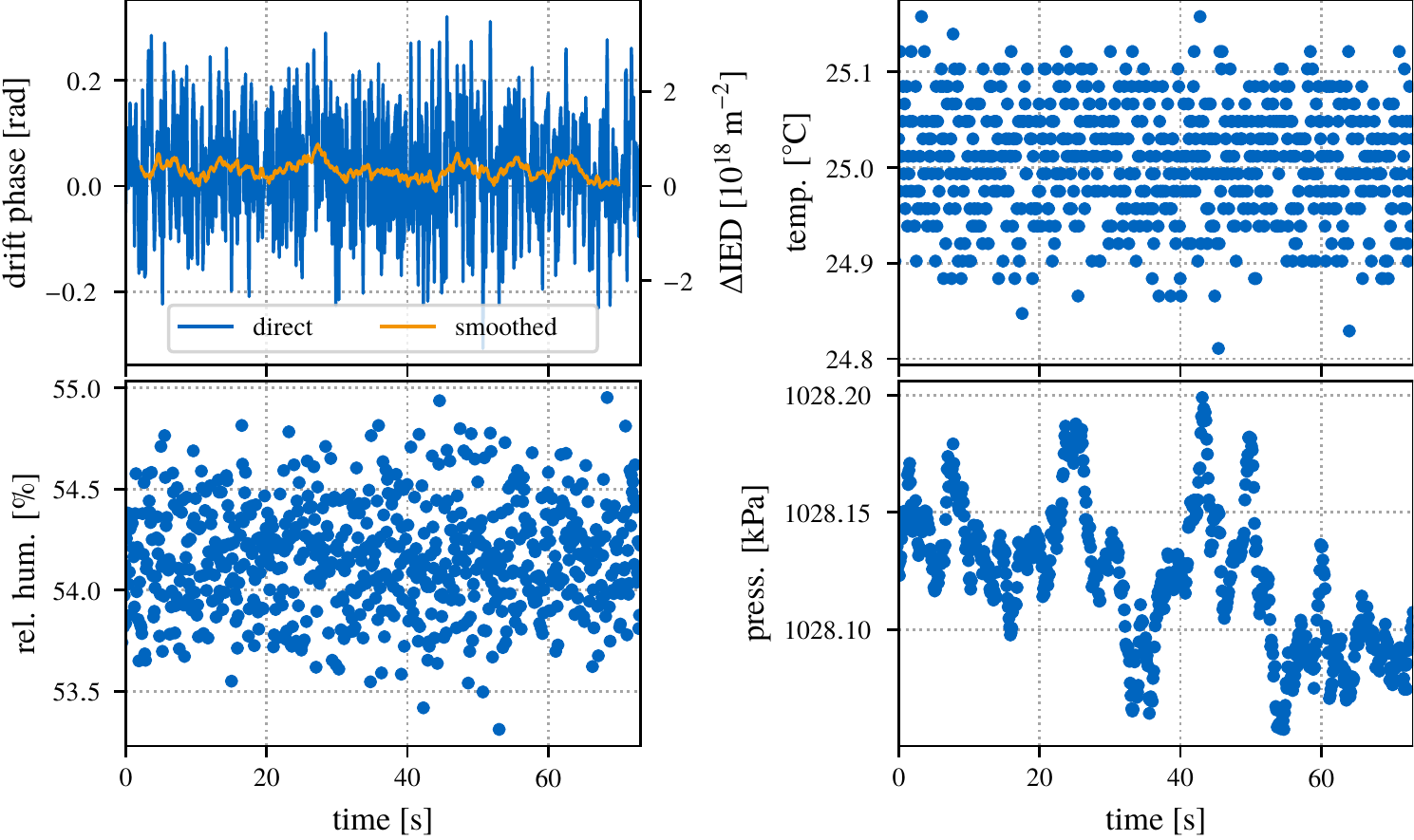}
  \caption{\label{fig:driftPhase} A drift phase measurement for W7-X shot \#20180927.14. The top left shows the modeled phase $\Phi_{\text{env}}$ with the direct calculation in blue. The orange line was smoothed using a 30~sample box averaging filter. The other three plots show the environmental parameters for comparison.}
\end{figure}

The tests conducted here also showed that the measurement of humidity can result in significant statistical noise on the density measurement. This is again indicated in \cref{fig:driftPhase} in the top left. The fluctuations are of very short time scale, which is to a large extent statistical noise in the humidity measurement. A simple approach is to use a simple 30 sample box-averaging filter, which can easily be implemented on an FPGA. Given the same hardware such a filter can already reduce the error on the direct model-phase (in blue), which was used for this paper, below \SI{2e18}{\m^{-2}} (orange). This is well below the control accuracy of the W7-X density control system and approaches the statistical noise of the IEDDI system itself\cite{Brunner2018}.

It is conceivable that a significant portion of the fluctuations is caused by air turbulence affecting the sensor itself, e.g. by changing the local temperature (which the SenseHAT uses to calculate the rel. humidity). To circumvent this issue, multiple measurements of the local parameters could be taken at different locations. This of course comes at a cost, but multiple sensors could be managed by a single RaspberryPi (or equivalent miniPC).

With an even higher budget the relatively cheap SenseHAT sensor, which is specified with a rel. humidity error of \SI{4}{\%}, could be exchanged for a more accurate sensor. Since the time constants of interest for the model are well above \SI{1}{\s}, a significantly slower sensor can be accepted.

Eventually the system developed here will be used at W7-X in the OP2 operation campaign where discharge lengths of up to \SI{15}{\minute} are envisaged. Due to the recent changes in the ITER diagnostic layout it should also be considered, whether this compensation scheme becomes mandatory for the ITER interferometers. 

\acknowledgments

The authors wish to explicitly thank T.~Akiyama and M.~van~Zeeland for the very fruitful discussions on humidity induced phase drifts.

\acknowledgement

\section*{Sources}

Much of the evaluation in this article was conducted using Python 3.7 in combination with the numpy, matplotlib and pandas libraries\cite{python,pandas,matplotlib,numpy}.

The version of the FPGA firmware this paper is based can be found under\\\mbox{\url{https://gitlab.mpcdf.mpg.de/kjbrunne/ieddi_fpgaware}}.

The git commit 750d4005bdf986b39ef05757c5cf3effd3f09d51 was used for the contents of this article. This also includes the simulation code. 

The codes for data evaluation and plot generation can be supplied on request. 


\bibliographystyle{JHEP}
\bibliography{papers}

\end{document}